# Under Water Waste Cleaning by Mobile Edge Computing and Intelligent Image Processing Based Robotic Fish

*Subhadeep Sahoo, Xiao Han Dong, Zi Qian Liu, Joydeep Sahoo*

*Abstract*— As water pollution is a serious threat to underwater resources, i.e., underwater plants and species, we focus on protecting the resources by cleaning the non-biodegradable waste from the water. The waste can be recycled for further usage. Here we design a robotic fish which mainly comprises optical biosensor, camera module, piston module, and wireless transceiver. By exploiting the LTE and 5G network architecture, the fish stores the information about the underwater waste in the nearest mobile edge computing server as well as in the centralized cloud server. Finally, when the fish clears the underwater waste, it offloads the captured image of the located object to the mobile edge computing server or sometimes to the cloud server for making a decision. The servers employ intelligent image processing technology and an adaptive learning process to make a decision. However, if the servers fail to make a decision, then the fish utilizes its optical biosensor. By this scheme, the time delay for clearing any water body is minimized and the waste collection capacity of the fish is maximized. This technique can effectively help the government or municipal personnel for making clean water without manual efforts.

## I. INTRODUCTION

Water is the integrated part of nature which contains huge resources. All living beings can survive if and only if there is a biological balance between the surface world and the underwater world. In the popular movie named as "Aquaman", we have seen that the clash between the creatures of the surface world and the underwater world can destroy nature. But since the industrial revolution in the late 17th century, the water is getting polluted day by day, which caused many underwater species to die. In recent years, water pollution is growing as a big issue and it is a serious threat to the underwater species. So, solving this issue is gaining much attention from many researchers as well as industries.

Advancement in the domain of information and communication technology (ICT) is playing a big role in our day to day life. So, integration of ICT with the natural resources can actively solve the open challenges. LTE network has been deployed all over the world and 5G will be deployed world wide by 2020, the features of 5G wireless communication, i.e., ultra-low latency, high-speed data rate, reliability, can come into play to suppress many gaps. In past decades, cloud computing has gained popularity as it solves the higher computational task by offloading them into the cloud server and minimizes the energy consumptions at the users end. However, in cloud computing the latency is a crucial challenge and this can be solved by introducing the mobile edge computing (MEC) technology [1]. In MEC, by exploiting the small cell based network architecture, the users can offload their task to the nearest MEC server, which can minimize the task overloading at the user end, consume less power, minimize the size of the end devices, also the latency is minimized.

According to the theme "Dive Deeper", in this work we consider that the underwater resources needs to be protected from the serious threat of water pollution. In many countries the underwater waste is separated and cleared by the human being. This cleaning method is inefficient and maybe the accuracy is also not enough. For a huge water body, it requires much effort and huge manual power, which is not cost efficient. Also, human being can face lot of communicable diseases.

In order to solve the aforementioned problem, we design robotic fishes to clear the waste from the water bodies. We also integrate the MEC and cloud computing technology with the robotic fishes to make a tradeoff between minimizing the size of the computational board inside the fishes, and optimizing the transmission delay and higher computational tasks. We utilize optical biosensors, intelligent image processing with adaptive learning to effectively discriminate the water waste. The rest of our idea molecule is organized as follows. Section II proposes the network model. Section III presents all the components of the robotic fish. Section IV formulates the problem. Section V illustrates the proposed solution. In section VI, we presents the advantages of the scheme. Finally, section VII concludes this molecule.

## II. NETWORK MODEL

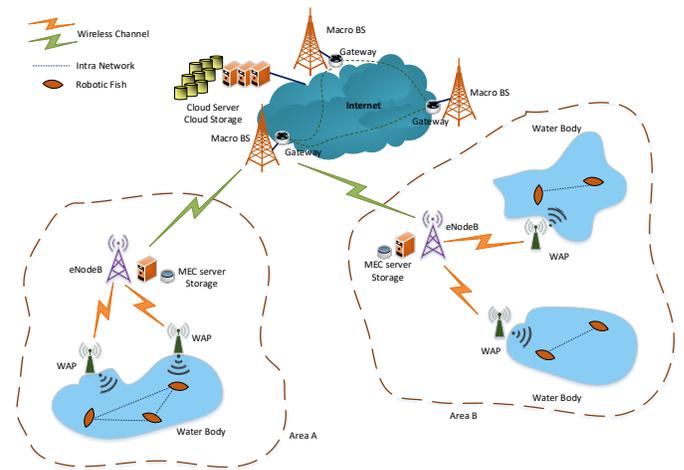

Fig. 1. System model.

We present our network model in Fig. (1). As the behavior of the water waste is nearly similar in an area, all the water bodies are classified and grouped into several particular areas. The wireless access points (WAPs) are placed near the water bodies. The WAP acts like the relay node or gateway between the fishes and the mobile edge computing (MEC) server. The robotic fishes themselves can create intra wireless network under the water, which is device to device (d2d) network. If any of the fishes moves out of the coverage area of all the WAPs, then the intra wireless network forwards the request to the WAPs. The WAPs and the robotic fishes consist of hybrid monopole-ring antenna [2], [3] which can achieve a wide bandwidth from 52.5 MHz to 162.5 MHz.

The MEC server is located at the eNodeB (small base station) inside a particular area. All these eNodeB are connected wirelessly with the macro base station (BS). The storage and computing capacity of the MEC servers is not high. The macro BS are the gateway to the internet, which is considered as the wide area network (WAN). The internet has the cloud server and storage. The cloud server has huge computing capacity and the storage capacity.

## III. COMPONENTS OF THE ROBOTIC FISH

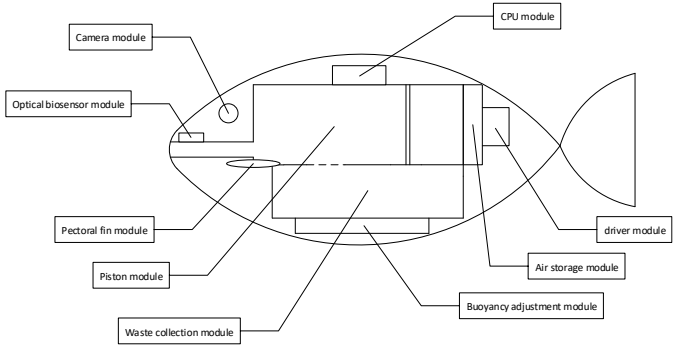

Fig. 2. Structure of the Robotic Fish

The fish body is partially modularly closed, the outer part is sealed with rubber, and the inner sealing shell has a circular cross section [4]. The processing technology is very simple, the processing cost is low, and the modular sealing can well avoid the water infiltration. All the components are presented below.

- **Optical Bio Sensor:** This sensor is mounted on the upper portion of the fish to discriminate between the biodegradable and non-biodegradable underwater waste. As discussed in [5], optical biosensors measure microbial growth of the waste in two ways, i.e., from emitting bacteria or by utilizing oxygen-sensitive colored indicator. In the oxygen-sensitive colored indicator, an optical fiber is covered with a fluorescent indicator and then microorganisms are immobilized over the optical fiber. If there is a presence of $O_2$, then it acts as a quencher and then attenuates the fluorescence intensity of the indicator. So, it can differentiate between the biodegradable and non-biodegradable waste to some degree. However, the main challenge of using this sensor is the time taken for a decision is very long and not quite convincible.

- **Camera Module:** The camera module is placed at the eye of the fish which captures the pictures of the objects located in front of the fish. This camera module works simultaneously with the optical biosensor in the first phase to associate the pictures and the decision from the optical biosensor. However, in the second phase, when the decision from the optical biosensor is not required, the camera module can work independently. The camera captures the pictures with a certain interval of time and sends the pictures to the CPU module of the fish for further computation.

- **CPU Module:** This CPU module is the decision maker and controller of the fish. This CPU controls the fish movement by instructing the driver module, also responsible for the functions of the piston module. In the first phase, it collects the decision and the associated picture from the optical biosensor and the camera respectively. The hybrid monopole-ring antenna is installed inside the CPU module to communicate between the robotic fishes and with the nearest WAP. The CPU module sends all these data to the nearest MEC server via WAP for storing them in the database of the MEC. In the second phase, after getting the image from the camera module, the CPU communicates with the MEC for the decision. Then, it instructs the piston module to collect or leave the object according to the decision. However, if there is no decision from the MEC, then the CPU module takes help from the optical biosensor to decide whether the located object is biodegradable or not. Then, sends the decision to the MEC server of the associated image for future usage.

- **Pectoral fin module:** According to the three-axis gyroscope in the mechanical fish body, the perception of horizontal and vertical declination, and the mechanical movement of the left and right pectoral fins, dynamically adjust the balance of the mechanical fish to ensure the waste collection process can be completed smoothly.

- **Piston module:** It uses the push-pull principle of the piston for waste absorption and transfers the acquired waste to the waste collection module.

- **Waste collection module:** It temporarily stores the waste obtained in the piston, this module occupies the maximum volume of the fish.

- **Buoyancy control module:** When the fish body inhales waste and water, it can sense the gravity and buoyancy of the fish body. By inhaling/discharging the water stored in the module, the buoyancy of the fish body is dynamically adjusted to maintain the stable position of the fish inside the water.

- **Air storage module:** It stores a part of the air that can be recycled to ensure the smooth movement of the piston.

## IV. PROBLEM FORMULATIONS

**Delay Calculation:**

According to the Shannon theory as presented in Eq. (1), the achievable data rate ($R_i$) of a wireless channel is inversely proportional to the distance ($d_i$). Here, $B_i$ is the bandwidth, $p_i$ is the transmitted signal power, $\sigma_i^2$ is the noise power, and $c$ is a constant value.

$$R_i = B_i \cdot \log_2 (1+SNR) = B_i \cdot \log_2 (1+\frac{p_i \cdot c}{\sigma_i^2 \cdot d_i^2}) \qquad (1)$$

In order to increase the data rate of the wireless channel between the fish and WAP, the fish always chooses to communicate with the nearest WAP placed on the surface. Assume, the distance between the fish and the nearest WAP is $d_1$, which can be varied. However, the distance ($d_2$) between the WAPs and the MEC is fixed.

The fish offloads the $k^{th}$ picture with the size of $D_k$ (in KB) to the assigned MEC server via the nearest WAP. To transmit the data to the MEC server, there will be transmission latency $T_k$. The transmission latency can be realized by Eq. (2). Here, $j$ is determined by the number of hops between the fish and the destination server. If it is computed in the local MEC server, then $j = 2$, otherwise $j$ differs for longer path.

$$T_k = D_k \cdot (\frac{1}{R_1} + \frac{1}{R_2} + ... + \frac{1}{R_n}) = D_k \cdot \sum_{j=1}^{n} \frac{1}{R_j} \quad (2)$$

When the transmitted image reaches at the server, it needs to be processed and the footprint of the image will be matched with the stored data in the database. As in Eq. (3) the processing also realizes latency $T_p$, which is inversely proportional to the computing capacity of the server $f_c$. Let assume that the equivalent computing capacity requirement for processing $D_k$ amount of data is $C_k$. By combining both the Eq. (2) and (3), the total latency ($T$) can be calculated according to the Eq. (4).

$$T_p = \frac{C_k}{f_c} \quad (3)$$

$$T = T_p + T_k \quad (4)$$

**Waste Capacity:**

With the continuous accumulation of waste and the inhalation and discharge of water during the collection of waste, the weight of the mechanical fish itself is constantly changing, and it is difficult to make the mechanical fish maintain a relatively stable state inside the water body. In order to solve this problem, the following analysis is proposed.

When the mechanical fish sneak into the water, to ensure the stability of the fish inside the water, the buoyancy of the mechanical fish should be equal to the gravity, which is realized in Eq. (5).

$$\rho \cdot V_w \cdot g + G = \rho \cdot g \cdot V_f \quad (5)$$

In Eq. (5), $\rho$ represents the density of water, $V_w$ is the volume of water inhaled from the fish, $g$ represents the acceleration of gravity, $G$ is the gravitational force of the robotic fish, $V_f$ represents the volume of the robotic fish.

When the mechanical fish collects waste once, the added gravity of the mechanical fish is expressed as the following Eq. (6), where $t$ represents the time required for piston 2 to move from the left end of container "a" to the right end, $v$ represents the velocity of the piston, $S$ represents the cross-sectional area of container "a", and $G_r$ represents the gravity of waste.

$$G_\Delta = \rho \cdot (t \cdot v \cdot S) \cdot g + G_r \quad (6)$$

In order to maintain the stability of the mechanical fish in the water, the buoyancy adjustment module needs to eject water with the same gravity. After ejection, the buoyancy and gravity should be equal as presented in Eq. (7).

$$\rho \cdot V_w \cdot g - \rho \cdot (t \cdot v \cdot S) \cdot G_r + G = \rho \cdot g \cdot V_f \quad (7)$$

Suppose the gravity of the remaining water in the buoyancy control module is $G_w$, which can be calculated by Eq. (8).

$$G_w = \rho \cdot V_w \cdot g - \rho \cdot (t \cdot v \cdot S) \cdot g - G_r \quad (8)$$

When $G_w$ is zero, make sure the following Eq. (9) must be followed.

$$G + G_r < \rho \cdot g \cdot V_f \quad (9)$$

V. PROPOSED SOLUTION

For the aforementioned problem, we propose MEC and intelligent image processing based robotic fish, which can efficiently clear the underwater non-biodegradable waste with lower delay. Our proposed solution has two phases. In the first phase, all the picture data and the decision associated with the pictures are collected by the fish and stored in all the MEC servers located in a particular area and in the cloud server. In the second stage, the fish actually clears the non-biodegradable waste from the water body. As the fish has low capacity of storage and computing capacity, it offloads the task to the nearest server. One more crucial challenge is if we only rely on the optical biosensor, then the time delay for a decision is very high. The second phase can go through three scenarios.

a) In the first scenario, the fish captures the picture and offloads it to the MEC server for making the decision. If the footprint of the transmitted image is presented in the MEC server, then MEC server processes the image according to the image processing algorithm (presented below) and sends back the decision to the fish.

b) In the second scenario, if the footprint of the transmitted image is not presented in the MEC server, the MEC server offloads this picture to the cloud server for making the decision. Similarly, if the footprint of the transmitted image is presented in the cloud server, then it processes the image according to the image processing algorithm (presented below) and sends the decision back to the fish via MEC server. If the decision is made, then the decision is sent to the fish and the decision with the associated image is stored in the MEC server.

c) In the final scenario, if the footprint of the transmitted image doesn't match with the database, both the cloud server and MEC cannot make the decision. Then the optical biosensor mounted in the fish takes the decision and the decision with the associated image is stored in the MEC and the cloud server for future usage. This causes higher time delay, so, it is employed only in the worst case scenario.

**5.1. Heuristic Algorithm:**

Here, we formally introduce the proposed solution in details by the following algorithms. The solution has two phases.

- **First phase: Data collection**

**Step 1:** Determine an underwater object as biodegradable or non-biodegradable by Optical biosensor of the fish.

**Step 2:** Capture the picture of the object by the camera module.

**Step 3:** Send both the decision and the associated picture the MEC server via nearest WAP.

**Step 4:** Store the data (i.e., picture and decision) in all the MEC servers which belong to the same area, as well as in the central cloud server.

- **Second phase: Waste clearance**

**Step 1:** Capture the image of the object in front of the fish and send it to MEC via nearest WAP.
**Step 2:** MEC processes the image, sends the decision back to the fish if the footprint is found in the MEC server and go to Step 5, otherwise, go to Step 3.
**Step 3:** MEC offloads the image to the centralized cloud server for further image processing. The cloud server sends back the decision if the footprint is found in the cloud server and MEC server forwards the decision to the fish, stores it in its database and go to Step 5, otherwise, go to Step 4.
**Step 4:** The fish determines the object as biodegradable or non-biodegradable by its Optical biosensor and sends the decision to both the MEC server and cloud server for updating their database.
**Step 5:** If the Eq. (9) satisfies, the fish acts on the waste according to the decision by calling Sub algorithm 1, and go to Step 1, otherwise, Stop.

- **Sub algorithm 1: Waste collection**

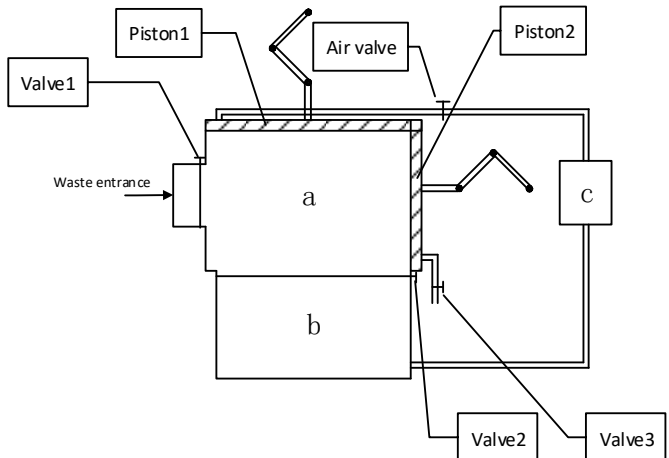

Fig. 3. Detailed Piston, Waste Collection and Air Storage Module.

As shown in Fig. 3, where "a" denotes the piston module, "b" denotes the waste collection module, and "c" denotes the air storage module. In the initial state, the valve 1 is opened, the valves 2 and 3 are closed, the piston 1 is at the upper end of "a", and the piston 2 is at the left end of "a". The detailed algorithm is presented below.

**Step 1:** After receiving the waste acquisition command, the arm of the piston 2 contracts, the piston 2 moves from the left end to the right end, closes the valve 1, opens the valve 3, the arm of the piston 2 is stretched, and the piston 2 moves from the right end to the left end, in the container "a" the water is ejected into the buoyancy adjustment module, and the valve 3 is closed.

**Step 2:** Open the air valve, the arm of the piston 2 contracts, the piston 2 moves from the left end to the right end, opens the valve 2, the arm of the piston 1 is stretched, the piston 1 moves from the upper end to the lower end, and the waste in the container a is pushed into the container "b", and the valve 2 is closed.

**Step 3:** The arm of the piston 1 contracts, the piston 1 moves from the lower end to the upper end, the arm of the piston 2 is stretched, the piston 2 moves from the right end to the left end, the air valve is closed, the valve 1 is opened, and Stop.

- **Sub algorithm 2: Image processing**

The image captured by the robotic fish in second phase is compared with the image data stored in the database of MEC server or cloud server. The image processing steps are given in Fig. 4. At first, the collected picture goes through the demosaicing and color balancing phase. In this phase, the image is reconstructed as a full color image from the incomplete color samples with a color filter array (CFA) and the colors are balanced. Then, the noise presented in the image is filtered out using DSP filter. The background is also filtered out and the object presented in the image is enhanced and it is stabilized. After that, the main enhanced object of the image is analyzed for the further decision. The context information (e.g., pixel) is extracted from the image. Then, the data is compared with the existing image data of the database for object recognition purpose. Finally, the image is tagged with a decision made by the server. For better performance, we can also utilize machine learning and artificial intelligence to recognize the objects.

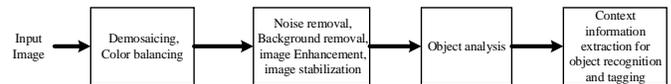

Fig. 4. Image Processing Steps

## VI. ADVANTAGES AND INTENDED CUSTOMERS

With the aim of reducing water pollution and protecting the underwater environment, the main advantages and the targeted customers of the proposed solution are as follows.

a) The proposed communication model and robotic fishes can efficiently separate the water waste and clear the non-biodegradable waste with low latency. As the capacity of waste collection of the fish is high, it can accumulate huge waste in a single run.
b) This system can minimize the human efforts in underwater cleaning, which further protects the human being from the affection of communicable diseases.
c) Several numbers of fishes can be re-utilized for different waterbody as the computation and storage depend on the MEC server, not on the fishes.
d) The intended customers of the proposed system are the government and the municipal personnel, who are in charge of clearing the water body (i.e., lakes, ponds, rivers, etc.) in a particular area.
e) This can facilitate the minimum cost of operation, implementation, and maintenance.

## VII. CONCLUSION

According to the theme "Dive Deeper", in this work, we introduced an innovative solution to segregate the underwater waste and remove the non-biodegradable waste, which can effectively save the underwater plants, species and contributes to the environmental balance. We integrated the latest LTE and 5G network architecture with the robotic fish to optimize the latency and increase the work efficiency. The artificial robotic fish can potentially remove underwater waste by utilizing intelligent image processing and adaptive learning policy. To minimize the performance delay, we employed MEC sever and cloud server in this idea due to their higher computation ability and huge storage capability. The potential customers of this solution can be the government and the municipal personnel. This solution can minimize manual work power and increases the accuracy of cleaning the underwater waste. The modern ICT is consolidated with each aspect of our life, so we can implement the ICT to look under the water and save the underwater ecology. As we are in the 5G era now, we can utilize its features implicitly and we hope this work is beneficial and cost effective for the customers.